\documentclass[a4paper, amsfonts, amssymb, amsmath, reprint, showkeys, footinbib, twoside,floatfix,longbibliography]{revtex4-1}

\usepackage{graphicx}%
\usepackage{dcolumn}%
\usepackage{bm}%
\usepackage{float}
\usepackage{xcolor}
\usepackage{mhchem}
\usepackage{gensymb}
\usepackage{verbatim}
\usepackage{newtxtext}
\usepackage[slantedGreek]{newtxmath}
\usepackage{amsmath}
\usepackage{longtable}
\usepackage[utf8]{inputenc}
\usepackage{lineno} 

\usepackage[T1]{fontenc}
\usepackage{hyphenat}
\usepackage{comment}
\usepackage[pdftex, bookmarks, pdffitwindow=false, pdfstartview=FitH, pdfdisplaydoctitle, colorlinks, plainpages=false, pdftitle={},pdfauthor={}, pdfpagelabels, hypertexnames, citecolor={blue!50!black},linkcolor={blue!50!black}, urlcolor={blue!50!black}, pdflang={en}, breaklinks]{hyperref}

\usepackage{siunitx}

\DeclareUnicodeCharacter{0308}{HERE!HERE!} 
\DeclareUnicodeCharacter{2212}{{\ensuremath{-}}}
\DeclareUnicodeCharacter{0308}{{\ensuremath{~}}}

\DeclareSIUnit\angstrom{\text {Å}}

\newcommand{\abs}[1]{\ensuremath{\lvert {#1}\rvert}}

\makeatletter
\renewcommand{\@seccntformat}[1]{}
\makeatother

\makeatletter
\def\@bibdataout@aps{
 \immediate\write\@bibdataout{
 @CONTROL{
   apsrev41Control, author="48",editor="1",pages="0",title="0",year="1"
 }}
 \if@filesw
  \immediate\write\@auxout{\string\citation{apsrev41Control}}
 \fi
}
\makeatother

\begin{document}

\preprint{}

%\linenumbers

\title{Approaching the Limit of Intrinsic Crystalline Thermal Insulation}

\author{Ruihuan Cheng\textsuperscript{1, $\dagger$}}
\author{Zhiqiang Cui\textsuperscript{1, $\dagger$}}
\author{Mani Jayaraman\textsuperscript{2}}
\author{Lincong Ji\textsuperscript{2}}
\author{Chen Wang\textsuperscript{3}}
\author{Zesheng Zeng\textsuperscript{1}}
\author{Petr Levinsky\textsuperscript{4}}
\author{Jiri Hejtmanek\textsuperscript{4}}
\author{Christophe Candolfi\textsuperscript{5}}
\author{Emmanuel Guilmeau\textsuperscript{6}}
\author{Xingchen Shen\textsuperscript{2, 6 $\ast$}}
\author{Yue Chen\textsuperscript{1,}}

\thanks{{Correspondence:}\\ \textcolor{blue}{\href{mailto:xingchen.shen@nwpu.edu.cn}{xingchen.shen@nwpu.edu.cn}};\\ \textcolor{blue}{\href{mailto:yuechen@hku.hk}{yuechen@hku.hk}}}

\affiliation{\textsuperscript{1}Department of Mechanical Engineering, The University of Hong Kong, Pokfulam Road, Hong Kong SAR, China}
\affiliation{\textsuperscript{2}MOE Key Laboratory of Material Physics and Chemistry under Extraordinary Conditions \& Shaanxi Provincial Key Laboratory of Condensed Matter Structure and Properties, School of Physical Science and Technology, Northwestern Polytechnical University, Xi’an 710072, China}
\affiliation{\textsuperscript{3}Institute for Advanced Study, Shenzhen University, Shenzhen 518060, China}
\affiliation{\textsuperscript{4}FZU - Institute of Physics of the Czech Academy of Sciences, Cukrovarnická 10/112, Prague 6 16200, Czech Republic}
\affiliation{\textsuperscript{5}Université de Lorraine, CNRS, IJL, F-54000 Nancy, France}
\affiliation{\textsuperscript{6}Laboratoire de Cristallographie et Sciences des Matériaux - UMR 6508, CNRS, Normandie Univ, ENSICAEN, UNICAEN, F-14000 Caen, France}
\affiliation{\textsuperscript{$\dagger$}These authors contributed equally to this work}

\date{\today}

\begin{abstract}

Crystalline materials with ultralow thermal conductivity ($\kappa$) are potential thermal barrier coatings or thermoelectrics, yet the discovery of ultralow-$\kappa$ materials remains inefficient due to the limitations of trial-and-error approaches. Herein, we propose a state-of-the-art high-throughput workflow that integrates universal machine learning interatomic potentials with high-fidelity phonon transport theories to accelerate the exploration of thermal insulators. Applying this approach, we identify dozens of crystalline materials with intrinsic room-temperature $\kappa$ values below 0.2 $\rm W m^{-1} K^{-1}$. Among them, we report and experimentally validate CsTlI$_4$, a record-breaking material with an ultralow $\kappa$ of 0.14 $\rm W m^{-1} K^{-1}$ at 300 K. Structural and bond analyses reveal that a hierarchical bonding framework, consisting of multi-coordinated Cs-I and antibonding Tl-I interactions, leads to weak chemical bonding and a soft lattice. These features reduce phonon group velocities, enhance phonon scattering, and induce strong vibrational mismatch between sublattices, collectively suppressing both particle-like phonon propagation and wave-like tunneling. Beyond this specific system, we establish physically interpretable descriptors based on interatomic force constants that correlate strongly with ultralow $\kappa$ and capture the role of bonding hierarchy and coordination environments in governing thermal transport. This work demonstrates a robust data-driven strategy for accelerating the discovery of thermal insulators and provides microscopic insight into how hierarchical bonding and strong anharmonicity cooperate to impede heat-carrying vibrations.

\end{abstract}

\maketitle
\section{Main}
Crystalline solids with intrinsically low thermal conductivity ($\kappa$) are of both fundamental interest and technological significance, as $\kappa$ is a key property of materials that governs the heat management in functional devices \cite{science1068609, jacs2c02017, tritt2005thermal}. In nonmetallic systems, heat is mainly transported by lattice vibrations in the form of phonons. Considering the efforts to minimize the heat flow and given the pivotal role of crystal structure and chemical bonding in governing thermal transport, extensive efforts have been devoted to exploring and designing materials that push the intrinsic limits of low lattice thermal conductivity  ($\kappa_{\rm L}$); various physical mechanisms have been proposed to explain such suppression \cite{scienceabk1176, adfm202108532, advs202417292}. 

Weak chemical bonding can give rise to large-amplitude vibrations of loosely bound atoms within an otherwise periodic crystalline lattice. This effect is particularly pronounced in structurally complex materials such as chalcogenides \cite{anie201605015, jacs3c04871}, clathrates \cite{PRL82779, CHAKOUMAKOS200080}, skutterudites \cite{science27252661325, D2TA02687B}, tetrahedrites \cite{PRL125085901, D4TA03316G}, and halide perovskites \cite{pnas1711744114, jacs1c11887}, where oversized cages often host guest atoms exhibiting rattling behavior. Such dynamics lead to strong anharmonicity and enhanced phonon scattering. Moreover, the weak bonding environment results in reduced phonon group velocities \cite{tritt2005thermal}, further impeding phonon transport and contributing to intrinsically low lattice thermal conductivity. Subsequent studies have revealed that a series of crystalline compounds, including CsAg$_5$Te$_3$ \cite{anie201605015}, CsSnI$_3$ \cite{pnas1711744114}, Cs$_2$SnI$_6$ \cite{acschemmater2c00084}, Cs$_3$Bi$_2$I$_9$ \cite{adfm202304607}, Cs$_3$Bi$_2$I$_6$Cl$_3$ \cite{acharyya2022glassy}, ternary Ag-Tl-I compounds \cite{zeng2024pushing, cheng2025strong}, and even the intermetallic compound Gd$_{117}$Co$_{56}$Sn$_{112}$ with exceptionally high structural complexity \cite{JACS2012Gd117Co56Sn112}, exhibit intrinsically $\kappa_{\rm L}$ values ranging from 0.2 to 0.4 $\rm W m^{-1} K^{-1}$, highlighting that ultralow $\kappa_{\rm L}$ can emerge from weak bonding, structural complexity, or a combination of both.

\begin{figure*}[htbp]
\includegraphics[width=1.0\linewidth]{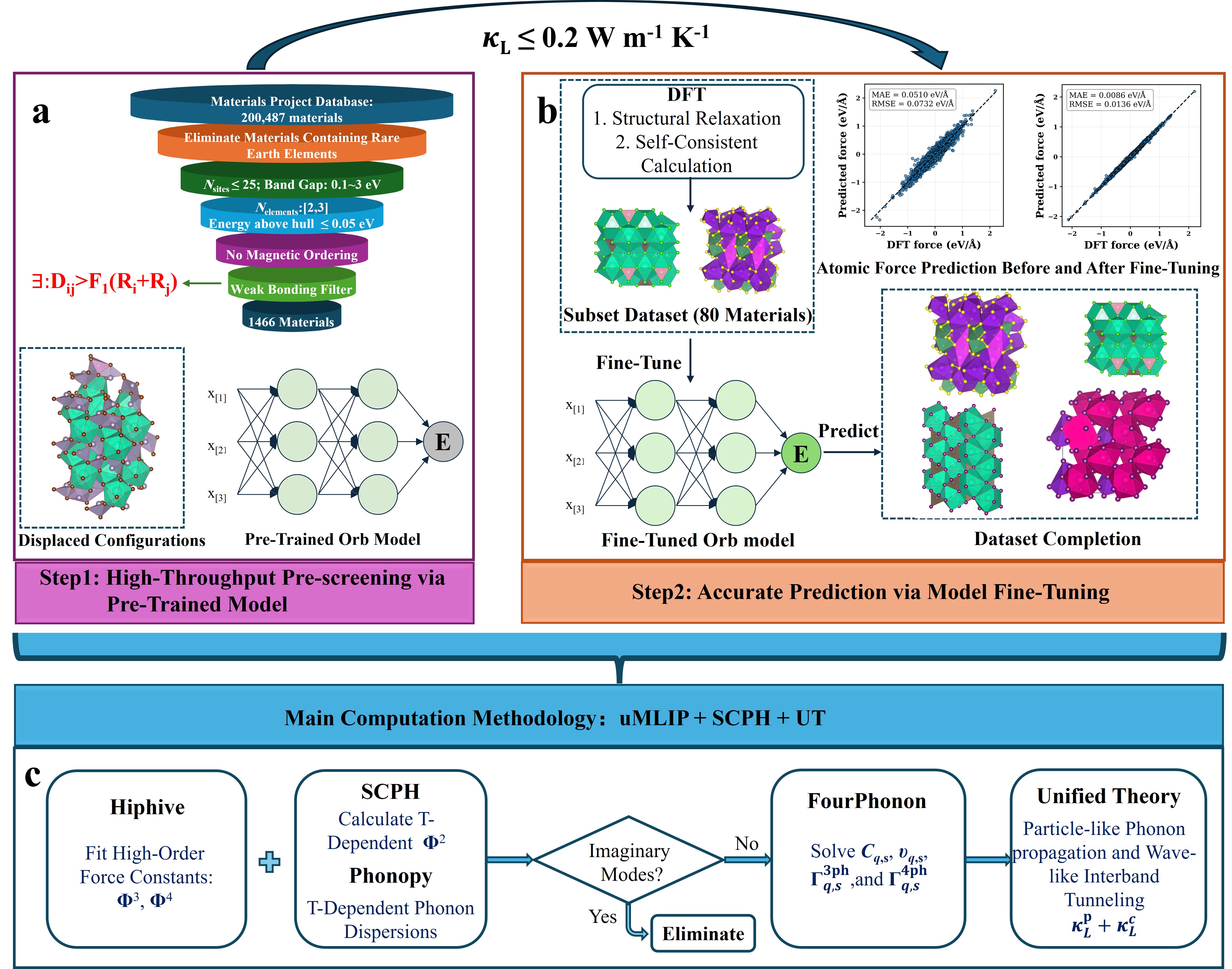}% Here is how to import EPS art
\caption{\label{fig:1} \textbf{Two-stage workflow for high-throughput discovery of ultralow lattice thermal conductivity materials.} \textbf{a}, Step 1: high-throughput prescreening of weakly bonded semiconductors from the Materials Project (MP) database, starting from 200,487 inorganic compounds. The screening criteria are applied sequentially as follows: exclusion of rare-earth-containing compounds, primitive-cell atom number $N_{\mathrm{atom}} \leq 25$, band gap in the range of 0.1–3.0 eV to focus on semiconductors relevant to thermoelectric and thermal-management applications, binary or ternary composition, energy above hull $\leq 0.05$ eV atom$^{-1}$, nonmagnetic character, and a weak-bond criterion for nearest-neighbor pairs, $D_{ij} > F_1(R_i+R_j)$ with $F_1 = 1.15$, where $D_{ij}$ is the interatomic distance and $R_i$ and $R_j$ are the covalent radii of atoms $i$ and $j$, respectively. This procedure yields 1466 candidate materials. A pretrained universal Orb machine-learning potential is then used to generate thermally displaced configurations at 300 K and predict the corresponding energies and forces, enabling rapid ML-accelerated screening of $\kappa_{\rm L}$ through the workflow shown in panel (\textbf{c}). \textbf{b}, Step 2: high-accuracy evaluation through material-specific model fine-tuning. From the initial screening results, 80 materials with $\kappa_{\rm L} \leq 0.2$ $\rm W m^{-1} K^{-1}$ are selected for further refinement. After structural relaxation, 200 thermally displaced configurations are generated for each material, among which 2\% (four configurations per material) are randomly selected for self-consistent DFT calculations to construct a compact reference force dataset. The pretrained Orb model is then fine-tuned using these data, yielding a force RMSE of 0.0136 eV\,\AA$^{-1}$ on the test set. The fine-tuned model is subsequently used to predict energies and forces for the remaining displaced configurations, and the same $\kappa_{\rm L}$ workflow in panel (\textbf{c}) is applied to obtain refined lattice thermal conductivities. \textbf{c}, Computational workflow for lattice thermal conductivity. High-order interatomic force constants are fitted using the Hiphive package; temperature-renormalized second-order force constants at 300 K are obtained using the self-consistent phonon (SCPH) method; phonon dispersions are calculated within the Phonopy package, and dynamically unstable structures with imaginary phonon modes are discarded. Specific heat $C_{qs}$, group velocity $v_{qs}$, and three-phonon and four-phonon scattering rates, ${\rm \Gamma}_{q, s}^{\rm{3 ph}}$ and ${\rm \Gamma}_{q, s}^{\rm{4 ph}}$, are then evaluated, and the total lattice thermal conductivity is finally obtained within unified theory as the sum of particle-like and wake-like contributions, $\kappa_{\rm L} = \kappa_{\rm L}^{\rm p} + \kappa_{\rm L}^{\rm c}$.}
\end{figure*}

Beyond weak bonding alone, the presence of hierarchical chemical bonding has also emerged as a critical mechanism in suppressing phonon-mediated thermal transport. For instance, in metal-organic frameworks (MOFs), the combination of weak metal-ligand interactions and covalently linked organic ligands forms a distinct hierarchy of bonding stiffness that strongly impedes phonon propagation \cite{acsami0c21220, islamov2023high}. Similarly, in superionic materials such as Ag$_8$SnSe$_6$ \cite{ren2023extreme}, KAg$_3$Se$_2$ \cite{rettie2021two, wang2023theoretical}, AgCrSe$_2$ \cite{li2018liquid, wang2023anisotropic}, and Cu$_4$TiSe$_4$ \cite{cheng2025atomic}, mobile ions dynamically hop or even diffuse within a rigid host lattice, creating a heterogeneous bonding network and inducing dynamic disorder that enhances phonon scattering. A notable example leveraging such a hierarchical bonding landscape is Bi$_4$O$_4$SeCl$_2$ \cite{scienceabh1619}, a 2D layered mixed-anion compound whose spatially ordered superlattice interfaces suppress acoustic phonon transport along the stacking direction, yielding an ultralow $\kappa_{\rm L}$ of 0.10 $\rm W m^{-1} K^{-1}$ at 300 K.

Despite recent progress in discovering materials with intrinsically low $\kappa_{\rm L}$, most advances have relied on a trial-and-error strategy. This approach is inherently inefficient and incurs high computational and experimental costs. As a result, developing accurate yet efficient computational screening techniques to rapidly identify thermal insulating crystals has attracted growing interest. Some studies have utilized empirical models to estimate $\kappa_{\rm L}$ for high-throughput screening \cite{C4EE03157A, acschemmater6b04179, jacs1c11887}; however, such methods often oversimplify phonon transport physics and fail to capture essential effects such as temperature-induced phonon energy renormalization \cite{PRB84180301, SCPH2014}, temperature-dependent anharmonic force constants \cite{PRL122075901}, four phonon scattering \cite{PRB93045202, PRB96161201}, and dual-channel transport mechanisms \cite{UT2019, isaeva2019modeling}, leading to significant deviations from experimental values. Alternatively, descriptor-based approaches have shown strong correlations with lattice thermal conductivity ($\kappa_{\rm L}$) by capturing key features including anharmonicity \cite{PRM4083809}, structural complexity \cite{zeng2024pushing}, average atomic mass \cite{adfm202108532}, sound velocity \cite{shen2025accelerated}, and antibonding states \cite{APR10106350227080}. These methods enable rapid identification of promising thermal insulators by leveraging physically meaningful descriptors that reflect phonon properties. Notably, several crystalline materials exhibiting intrinsically ultralow $\kappa_{\rm L}$ were discovered based on these principles, including AgTlI$_2$ (0.26 $\rm W m^{-1} K^{-1}$) \cite{zeng2024pushing}, AgTl$_2$I$_3$ (0.21 $\rm W m^{-1} K^{-1}$) \cite{cheng2025strong}, and CsAg$_2$I$_3$ (0.16 $\rm W m^{-1} K^{-1}$) \cite{shen2025accelerated}, demonstrating the efficacy of descriptor-guided strategies in revealing materials with strongly suppressed thermal transport. Nonetheless, accurate predictions of thermal properties remain limited by the trade-off between computational cost and precision: first-principles calculations offer high fidelity but are too expensive for large-scale screening \cite{zhu2024high, LEE2025101688}. Fortunately, the recent emergence of universal machine learning interatomic potentials (uMLIPs), such as CHGNet \cite{deng2023chgnet}, MACE \cite{JCP50297006}, AlphaNet \cite{yin2025alphanet}, eSEN \cite{fu2025learning}, and Orb \cite{rhodes2025orb}, which approach DFT-level accuracy while being orders of magnitude faster, offers a viable pathway toward efficient and scalable prediction of a wide range of materials properties. They have already been applied to compute thermodynamic and dynamic stability, harmonic phonons, and interatomic force constants, as well as to characterize the strength of anharmonicity and the hierarchy of thermal transport in simple binary systems \cite{pota2024thermal} and cubic or tetragonal crystals \cite{li2025high}. These advances establish a promising foundation for rapid prediction of thermal properties, yet current implementations are still constrained by limited transferability, simplified phonon descriptions, and insufficient treatment of higher-order anharmonic effects, particularly for the large-scale identification of materials with ultralow $\kappa_{\rm L}$, where a broadly applicable and physically rigorous screening framework is still lacking.

\begin{figure*}[htbp]
\includegraphics[width=1.0\linewidth]{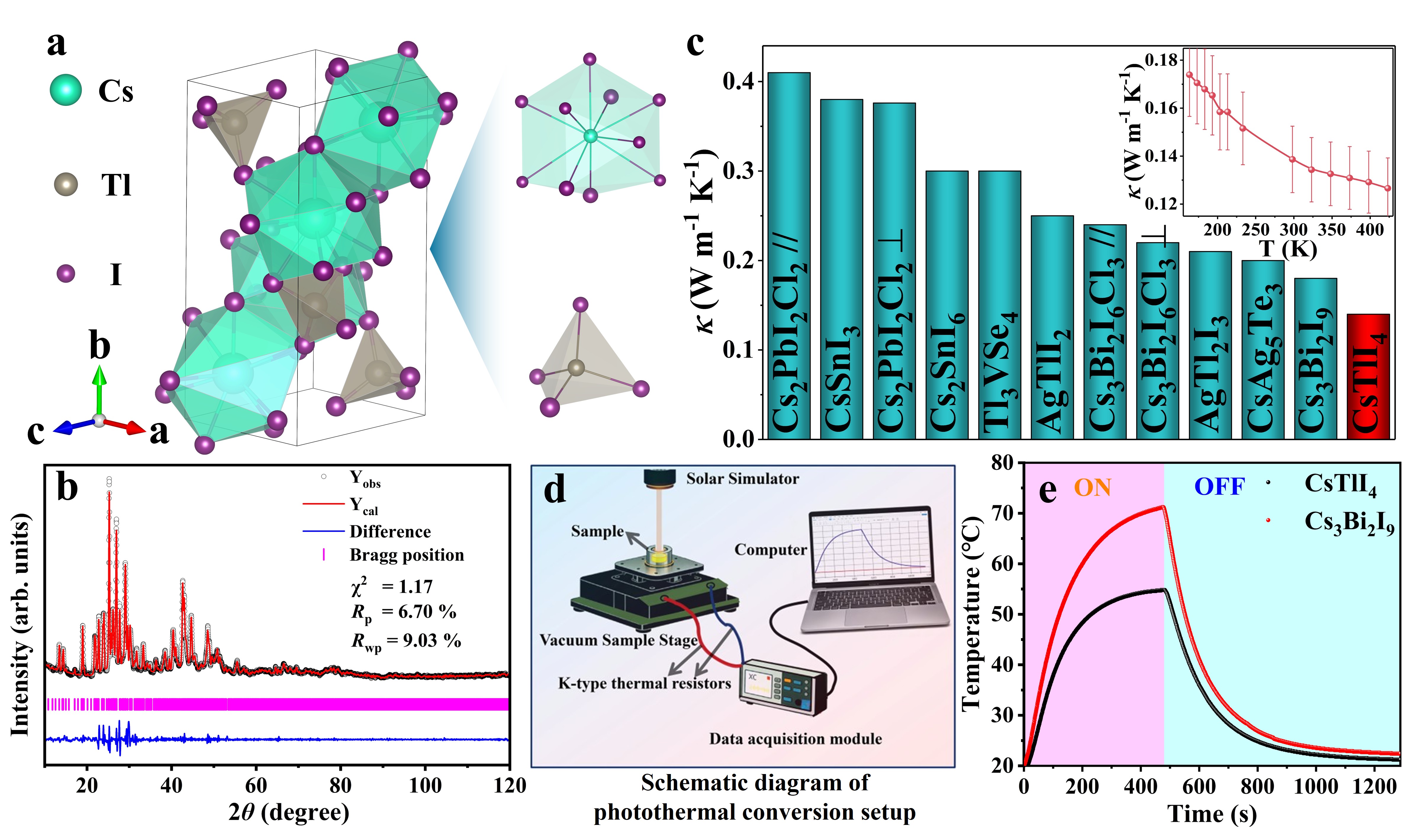}% Here is how to import EPS art
\caption{\label{fig:2} \textbf{Structural characterization and ultralow thermal conductivity of CsTlI$_4$.} \textbf{a}, The crystal structure of CsTlI$_4$, including the distinct coordination environments of Cs and Tl atoms and their atomic thermal vibration ellipsoids, was visualized using VESTA \cite{VESTA}. \textbf{b}, Rietveld refinement pattern for the sintered CsTlI$_4$ pellet, with XRD data collected at room temperature over a 2$\rm \theta$ range of 10 to 120$\degree$. The observed data, calculated pattern, difference plot, and Bragg peak positions are represented as black open circles, red solid line, blue line, and vertical pink tick marks, respectively. \textbf{c}, A comparison of the thermal conductivity ($\kappa$) of CsTlI$_4$ at room temperature with that of other representative ultralow-$\kappa$ materials \cite{jacs0c08044, pnas1711744114, acschemmater2c00084, scienceaar8072, zeng2024pushing, acharyya2022glassy, cheng2025strong, anie201605015, adfm202304607}. The inset shows the experimental thermal conductivity of CsTlI$_4$ measured from 163 K to 423 K. The experimental error represents the sum of instrument error together with the error in determining $C_p$ arising from the simplification based on the Dulong-Petit approximation. \textbf{d}, Schematic of the photothermal conversion setup. The system consists of a solar simulator (BETICAL, HDL-II DL, China), a vacuum sample stage, two K-type thermal resistors, a data acquisition module, and a computer. \textbf{e}, Temperature of the CsTlI$_4$ and Cs$_3$Bi$_2$I$_9$ pellets during irradiation ON/OFF cycles.}
\end{figure*}

\begin{figure*}[htbp]
\includegraphics[width=1.0\linewidth]{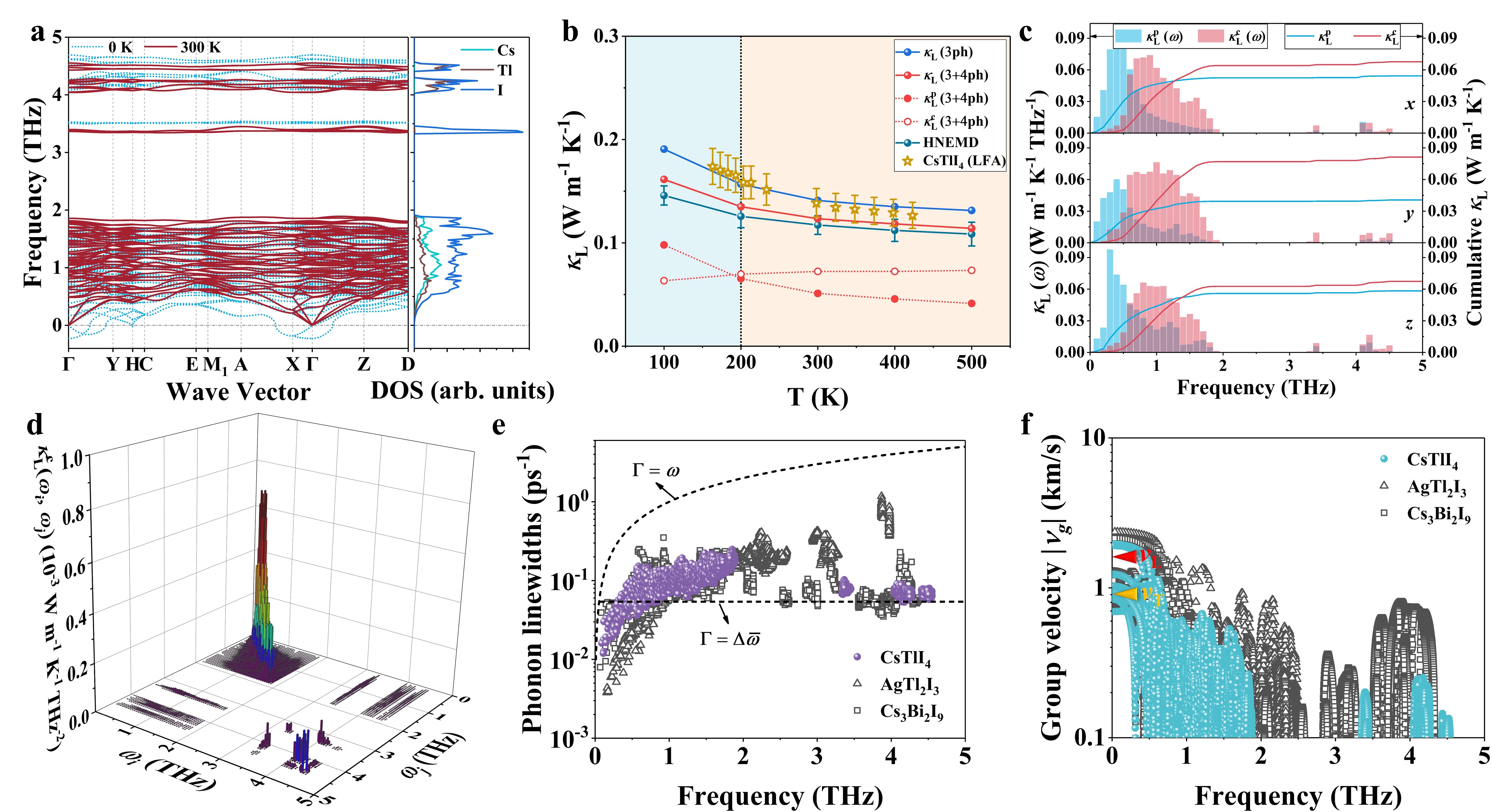}% Here is how to import EPS art
\caption{\label{fig:3} \textbf{Dual-channel thermal transport analysis reveals ultralow $\kappa_{\rm L}$ in CsTlI$_4$.} \textbf{a}, Phonon dispersions and atom-projected phonon density of states (DOS) of CsTlI$_4$ calculated at 300 K using the self-consistent phonon (SCPH) method \cite{PRB1572, SCPH2014}. For comparison, harmonic phonon dispersions computed at 0 K using the finite displacement method (FDM) \cite{TOGO20151} are overlaid as cyan-dotted lines. \textbf{b}, Average lattice thermal conductivity ($\kappa_{\rm L}$) of CsTlI$_4$ predicted using the unified thermal transport theory with only three-phonon scattering (3ph) and combined three- and four-phonon scattering (3+4ph), alongside results from homogeneous nonequilibrium molecular dynamics (HNEMD) simulations. Experimental data are included for validation and comparison. \textbf{c},\textbf{d} The spectral and cumulative particle-like phonon propagation ($\kappa^{\rm p}_{\rm L}$) and wave-like interband tunneling ($\kappa^{\rm c}_{\rm L}$) contributions to the lattice thermal conductivity of CsTlI$_4$ at 300 K along different directions \textbf{(c)} and three-dimensional mapping of $\kappa^{\rm c}_{\rm L}$ along the \textit{x} axis \textbf{(d)}. \textbf{e},\textbf{f} Phonon linewidths \textbf{(e)} and group velocities \textbf{(f)} of CsTlI$_4$ at 300 K, and a comparison with other ultralow-$\kappa$ materials AgTl$_2$I$_3$ \cite{cheng2025strong} and Cs$_3$Bi$_2$I$_9$. The curved dashed line marks $\rm \Gamma$ = $\omega$, while the horizontal dashed line in panel \textbf{e} denotes the Wigner limit (average interband spacing $\Delta\bar{\omega}$) \cite{PRX12041011}. The experimentally measured longitudinal ($v_{\rm L}$) and transverse ($v_{\rm T}$) sound velocities in panel \textbf{f} of CsTlI$_4$ are included for direct comparison with theoretical results.}
\end{figure*}

To meet these challenges, we propose an advanced high-throughput computational framework that combines the fine-tuned Orb-v3 \cite{rhodes2025orb} with temperature-renormalized phonon energy \cite{SCPH2014}, temperature-dependent anharmonic force constants, three- and four-phonon scattering processes \cite{Lindsay_2008, PRB93045202}, and the unified thermal transport theory \cite{UT2019}. This workflow enables both accurate and efficient prediction of $\kappa_{\rm L}$ and was carefully validated against experimental benchmarks. Leveraging this framework along with a curated database of structurally and chemically diverse crystalline compounds, we screened a large number of candidate materials and identified CsTlI$_4$ as a dynamically stable crystal exhibiting a theoretical $\kappa_{\rm L}$ of approximately 0.12 $\rm W m^{-1} K^{-1}$ at 300 K, which was subsequently confirmed by experimental measurements. Further chemical bonding and lattice dynamics analyses reveal that the ultralow $\kappa_{\rm L}$ in CsTlI$_4$ arises from the presence of weak Cs–I interactions associated with high coordination numbers, as well as a hierarchical bonding structure formed by Cs-I and Tl-I sublattices. These features result in strong lattice anharmonicity in soft lattices and vibrational mismatch, which collectively induce strong phonon scattering and low group velocities that suppress both particle-like and wave-like thermal transport channels. Building on these insights, we propose a set of structural and chemical bonding descriptors, including coordination number, bonding weakness, and bonding mismatch, that correlate strongly with low $\kappa_{\rm L}$. These descriptors are broadly observed in many low-$\kappa_{\rm L}$ compounds and provide robust physical criteria for rapidly identifying new thermal insulators in future materials discovery. 

\section{Results}
\subsection{High-Throughput Screening}

As illustrated in Fig. \ref{fig:1}, we employed a two-stage high-throughput computational framework to identify crystalline materials with ultralow $\kappa_{\rm L}$. The first stage combines weak-bond descriptors with a pretrained Orb model to rapidly narrow the search space and estimate $\kappa_{\rm L}$. The second stage applies material-specific fine-tuning to the most promising ultralow-$\kappa_{\rm L}$ candidates, enabling improved quantitative accuracy.

The first-stage screening is motivated by the well-established correlation between weak bonding and suppressed thermal transport through reduced phonon group velocities and enhanced anharmonicity \cite{jana2017intrinsic, jacs9b10551}. After applying filters on thermodynamic stability, band gap, cell complexity, chemical composition, and magnetic character, the initial Materials Project database was reduced to 1466 weakly bonded semiconductors (Fig. \ref{fig:1}a). This prescreening substantially lowers the computational cost while enriching the candidate pool with materials more likely to exhibit soft phonons, low group velocities, and strong anharmonicity. Among them, 791 compounds remain dynamically stable at 300 K, corresponding to a retention ratio of approximately 54\%, indicating that weak-bonding filters favorable for suppressing heat transport often lie near the boundary of lattice stability. For these 791 dynamically stable candidates, the pretrained Orb workflow yields a $\kappa_{\rm L}$ distribution strongly enriched in the low-$\kappa_{\rm L}$ regime. As summarized in the \textbf{Supplementary Form I}, 80 materials exhibit $\kappa_{\rm L}$ below 0.2 $\rm W m^{-1} K^{-1}$, 154 fall below 0.3 $\rm W m^{-1} K^{-1}$, 279 below 0.5 $\rm W m^{-1} K^{-1}$, and 414 below 1.0 $\rm W m^{-1} K^{-1}$. These results demonstrate that the weak-bond prescreening combined with pretrained Orb efficiently concentrates the search onto a relatively small subset of highly promising thermal insulators.

For the 80 candidates with predicted $\kappa_{\rm L} < 0.2$ $\rm W m^{-1} K^{-1}$, material-specific fine-tuning substantially improves force accuracy. As shown in Fig.\ref{fig:1}b, the root-mean-square error of atomic forces decreases from 0.0732 to 0.0136 eV Å$^{-1}$, representing an approximately fivefold improvement. This result shows that the pretrained model already provides a useful starting point, while a small amount of reference data from first-principles calculations is sufficient to significantly improve the description of the local potential energy surface. The improved force description leads to refined transport predictions. Among the 80 second-stage candidates, 72 remain dynamically stable after refinement (see \textbf{Supplementary Form II}). As summarized in Table S2, 34 materials still exhibit $\kappa_{\rm L}$ below 0.2 $\rm W m^{-1} K^{-1}$ and 59 remain below 0.3 $\rm W m^{-1} K^{-1}$. Notably, some compounds show exceptionally low values, including CsTlI$_4$, Ga$_2$SnI$_8$, CsHg$_2$Br$_5$, KI$_4$, GaTeBr$_7$, CsInI$_4$, and Tl$_3$AsSe$_3$, all with $\kappa_{\rm L}$ below 0.15 $\rm W m^{-1} K^{-1}$, among which CsTlI$_4$ shows the lowest predicted value of 0.12 $\rm W m^{-1} K^{-1}$. Interestingly, CsTlI$_4$ \cite{ThieleRotterZimmermann1986}, CsHg$_2$Br$_5$ \cite{ZNK_1977}, and Tl$_3$AsSe$_3$ \cite{Feichtner72} have already been synthesized experimentally, although their thermal transport properties remain largely unexplored. Overall, the refined $\kappa_{\rm L}$ are slightly higher than the prescreened values. This shift can be attributed to the lattice contraction after structural relaxation and the more accurate description of interatomic interactions provided by material-specific fine-tuning. Further validation using 19 experimentally reported low-$\kappa_{\rm L}$ compounds \cite{scienceaar8072, pnas1711744114, jacs7b02399, JAP12756037, jacs9b10551, anie201511737, jana2017intrinsic, hu2021high, WANG202167, adfm202304607, acharyya2022glassy, advs202400258, acschemmater2c00084, jacs0c08044, anie201605015, DGerlich_1982, cheng2025strong, pnas2521353123} shows good agreement with measurements (Table S1 and Fig. S1), yielding a mean absolute error of 0.0348 W m$^{-1}$ K$^{-1}$. This accuracy, together with the broad chemical coverage of the benchmark set, supports the predictive capability of the present two-stage workflow.

\subsection{Crystal Structure and Thermal Conductivity}

\begin{figure*}[htbp]
\includegraphics[width=1.0\linewidth]{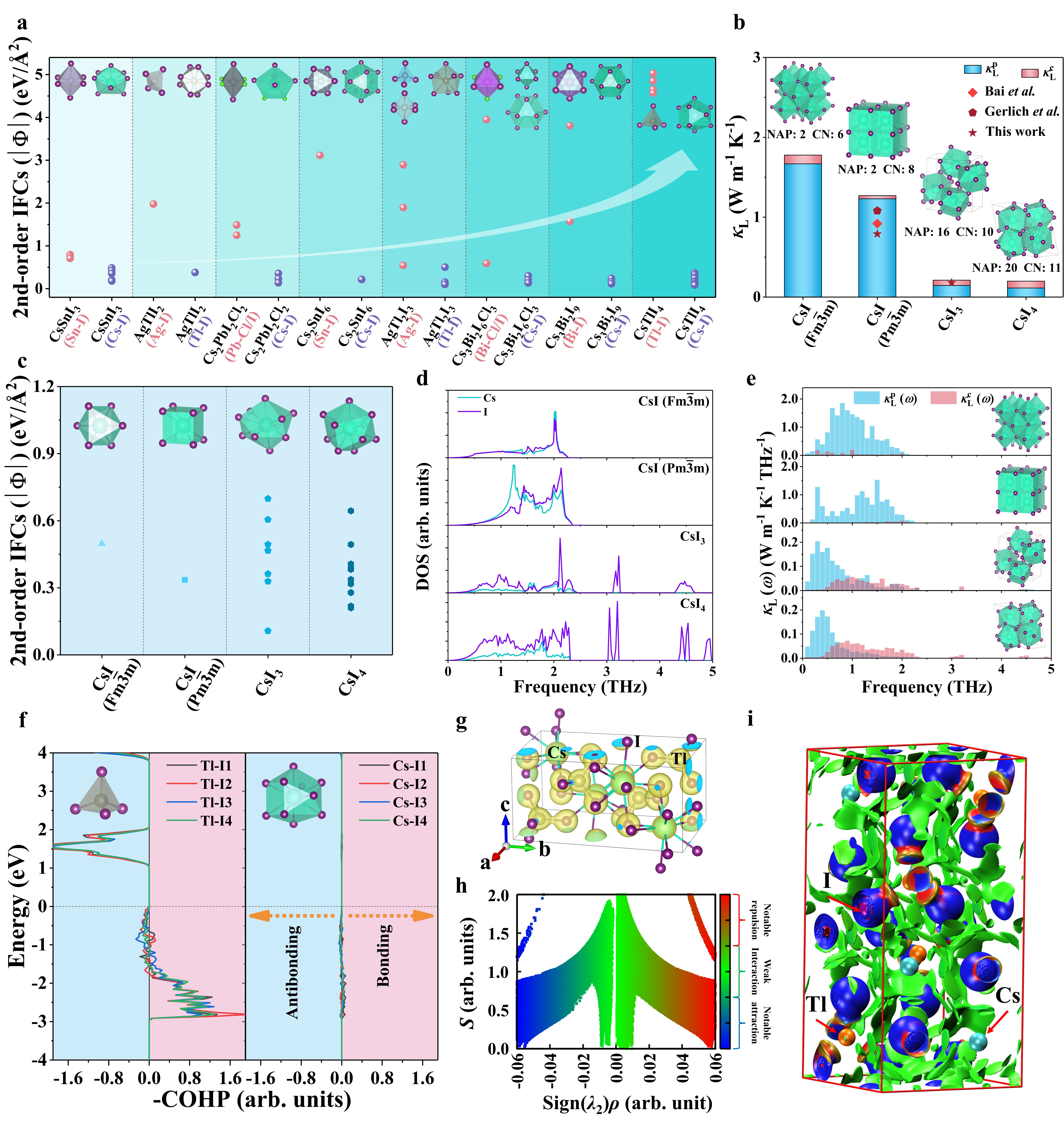}% Here is how to import EPS art
\caption{\label{fig:4} \textbf{Hierarchical bonding and phonon transport in CsTlI$_4$ and Cs–I systems.} \textbf{a}, Harmonic interatomic force constants (IFCs) for nearest-neighbor atom pairs in CsTlI$_4$, benchmarked against representative ultralow-$\kappa$ materials. Materials are arranged in order of increasing stiffness anisotropy, as represented by the ratio of the maximum to the minimum IFC values. \textbf{b}, The calculated lattice thermal conductivities of CsI (Fm$\bar{3}$m), CsI (Pm$\bar{3}$m), CsI$_3$, and CsI$_4$ at 300 K, including particle-like phonon propagation ($\kappa^{\rm p}_{\rm L}$) and wave-like interband tunneling ($\kappa^{\rm c}_{\rm L}$). The solid symbols represent experimental measurement data \cite{nwaf544, DGerlich_1982}. NAP: The number of atoms in the primitive cell. \textbf{c-e}, Harmonic IFCs for nearest-neighbor atom pairs \textbf{(c)}, atom-projected phonon DOS \textbf{(d)}, and spectral $\kappa^{\rm p}_{\rm L}$  and $\kappa^{\rm c}_{\rm L}$ \textbf{(e)} of CsI (Fm$\bar{3}$m), CsI (Pm$\bar{3}$m), CsI$_3$, and CsI$_4$. \textbf{f}, Crystal orbital Hamilton population (-COHP) analysis of Tl-I and Cs-I interactions in CsTlI$_4$. Positive and negative values of -COHP indicate bonding and anti-bonding states, respectively. The energy is shifted to the Fermi level at 0 eV. \textbf{g}, The total charge density of CsTlI$_4$, where an isosurface of the charge density has been visualized at 0.05 \textit{e}/Bohr$^3$. \textbf{h}, Non-covalent interaction analysis of CsTlI$_4$ based on the reduced density gradient (RDG) $S$ versus sign($\lambda_2$)$\rho$. Negative and positive sign($\lambda_2$)$\rho$ denote attractive and repulsive interactions, respectively, and weak interactions are pervasive in the vicinity of sign($\lambda_2$)$\rho$ = 0. \textbf{i}, Isosurface map of sign($\lambda_2$)$\rho$ with a standard coloring method. The blue, green, and red areas represent covalent, van der Waals, and repulsive interactions, respectively.}
\end{figure*}

To further validate the predictive accuracy of our high-throughput framework and understand the microscopic origins of suppressed thermal transport, we selected CsTlI$_4$ for detailed experimental and theoretical investigation. As shown in Fig. \ref{fig:2}a, CsTlI$_4$ crystallizes in a monoclinic structure (space group: $P2_1/c$) comprising Cs-centered polyhedral coordinated by nine nearest iodine ions and isolated [TlI$_4$]$^-$ tetrahedra. The bond lengths of the covalent Tl-I pairs range from 2.71 to 2.74 \AA. The Cs-I distances vary from 3.95 to 4.37 \AA, which are close to or slightly larger than the sum of the ionic radii of Cs$^+$ (1.70 \AA) and I$^-$ (2.20 \AA), suggesting a pronounced ionic bonding nature for the Cs-I interactions. This ionic bonding character favors weak chemical bonds, leading to large atomic displacement parameters for Cs and I, as listed in Table S3. Meanwhile, the isolated environment of I within the [TlI$_4$]$^-$ tetrahedra also provides a high degree of freedom for I ions. The combination of the ionic Cs-I bonding nature and the isolated iodine environment within the [TlI$_4$]$^-$ tetrahedra together contributes to strengthening thermal vibrations of the Cs and I atoms and lattice anharmonicity in CsTlI$_4$ (Tables S3 and S8). To validate the predicted ultralow $\kappa_{\rm L}$ of CsTlI$_4$, we synthesized the compound and performed detailed structural and thermal characterizations. Rietveld refinement of the pellet X-ray diffraction data (Fig. \ref{fig:2}b) at 300 K confirms that CsTlI$_4$ crystallizes in the expected structure with high phase purity, and the refined lattice parameters agree well with theoretical predictions (Tables S3 and S8), collectively providing a reliable basis for subsequent analysis. The thermal conductivity measurements shown in Figs. \ref{fig:2}c and S3 reveal that the CsTlI$_4$ sample exhibits a record-low $\kappa_{\rm L}$ of 0.14 $\rm W m^{-1} K^{-1}$ at room temperature. This value sets a new record among experimentally synthesized thermal insulators, outperforming Cs$_3$Bi$_2$I$_9$, which exhibits an ultralow $\kappa_{\rm L}$ of $\sim$0.20 $\rm W m^{-1} K^{-1}$ \cite{adfm202304607}. The inset further reveals a monotonic decrease in $\kappa_{\rm L}$ from 0.17 to 0.12 $\rm W m^{-1} K^{-1}$ between 163 and 423 K, which is consistent with enhanced phonon–phonon scattering at elevated temperatures, while CsTlI$_4$ remains the lowest $\kappa_{\rm L}$ over the full temperature range. To more intuitively demonstrate the thermal insulation performance, photo-thermal conversion measurements were conducted using a home-made setup, as illustrated in Fig. \ref{fig:2}d. Pellet samples of CsTlI$_4$ and Cs$_3$Bi$_2$I$_9$ (diameter: 10 mm; thickness: 1.5 mm) were placed on the vacuum stage, and the chamber was evacuated to $10^{-2}$ bar. Prior to illumination, all samples were stabilized at \SI{20}{\degreeCelsius}. The measurements were performed under a light power density of 10,000 W m$^{-2}$ with a wavelength range of 400–1100 nm. The temperature evolution of the two samples during light ON/OFF cycles is shown in Fig. \ref{fig:2}e. Under identical irradiation conditions, the bottom surface temperature of CsTlI$_4$ increases much more slowly than that of Cs$_3$Bi$_2$I$_9$. Upon switching off the illumination, CsTlI$_4$ also exhibits a slower cooling response. These results indicate reduced thermal transport through the CsTlI$_4$ sample, reflecting its lower $\kappa_{\rm L}$. This direct observation is consistent with the $\kappa_{\rm L}$ derived from thermal diffusivity measurements, further confirming the record-breaking thermal insulation performance of CsTlI$_4$ and motivating a deeper investigation of the underlying phonon transport mechanisms.

\subsection{Dual-Channel Thermal Transport}

We carried out detailed phonon calculations and dual-channel transport analysis, as summarized in Fig. \ref{fig:3}. The phonon dispersions and atomic projected density of states (Fig. \ref{fig:3}a) were calculated at 300 K using the self-consistent phonon (SCPH) method. Notably, the harmonic phonon dispersions computed at 0 K exhibits imaginary frequencies near the $\rm \Gamma$ point, suggesting that the harmonic approximation is insufficient to accurately describe the lattice dynamics of CsTlI$_4$. This result underscores the necessity of incorporating higher-order anharmonic effects in CsTlI$_4$ that exhibits harmonic instability at 0 K but is stabilized at finite temperatures. Upon phonon renormalization, the dispersions show that most phonon modes are concentrated below 2 THz, primarily contributed by the heavy Cs and Tl atoms. The iodine atoms participate across the full frequency range. To quantify the thermal transport of CsTlI$_4$, we employed both HNEMD simulations and the unified thermal transport theory \cite{UT2019}, integrating temperature-renormalized force constants (Fig. S4) and three- and four-phonon scattering processes (Fig S5). The latter also enables decomposition of the total lattice thermal conductivity into particle-like contributions from phonon propagation ($\kappa^{\rm p}_{\rm L}$) and wave-like contributions arising from interband tunneling coherence ($\kappa^{\rm c}_{\rm L}$). As shown in Fig. \ref{fig:3}b, both theoretical approaches yield results in close agreement with experimental measurements, and the consistency between HNEMD and unified thermal transport theory further suggests that phonon scattering processes beyond four-phonon interactions have a negligible impact in CsTlI$_4$. At 200 K, these two channels contribute nearly equally to the total $\kappa_{\rm L}$. As the temperature increases, the role of $\kappa^{\rm c}_{\rm L}$ becomes increasingly significant. This trend suggests that enhanced phonon scattering at elevated temperatures more strongly suppresses the propagation channel, while leaving the coherent tunneling channel relatively less affected. A distinct phonon gap above 2 THz plays a critical role in shaping the thermal transport behavior of CsTlI$_4$. This gap effectively suppresses interband tunneling processes, thereby limiting the contribution from wave-like coherent transport ($\kappa^{\rm c}_{\rm L}$). Spectral decomposition of the lattice thermal conductivity reveals that thermal transport is predominantly governed by low-frequency phonons below the gap as shown in Fig. \ref{fig:3}c. Modes above 2 THz contribute minutely due to their flattened dispersions, negligible group velocities, and diminished coherence. As illustrated in the 3D mapping along the \textit{x}-axis (Figs. \ref{fig:3}d and S6), the wave-like tunneling contribution ($\kappa^{\rm c}_{\rm L}$) primarily originates from nearly degenerate low-frequency phonons, which retain sufficient phase coherence to support interband tunneling. The phonon linewidths and group velocities in Fig. \ref{fig:3}e–f further support this picture. CsTlI$_4$ exhibits large phonon linewidths at low frequencies, which are comparable to Cs$_3$Bi$_2$I$_9$ and significantly larger than AgTl$_2$I$_3$, reflecting strong phonon scattering associated with dense low-frequency modes. Linewidths drop sharply above the phonon gap, further suppressing coherence. The group velocities are consistently smaller across the spectrum compared to Cs$_3$Bi$_2$I$_9$ and AgTl$_2$I$_3$, resulting in the simultaneous suppression of dual-channel transport and ultimately yielding the record-low $\kappa_{\rm L}$ observed.

\subsection{Chemical Bonding Analysis}

\begin{figure*}[htbp]
\includegraphics[width=1.0\linewidth]{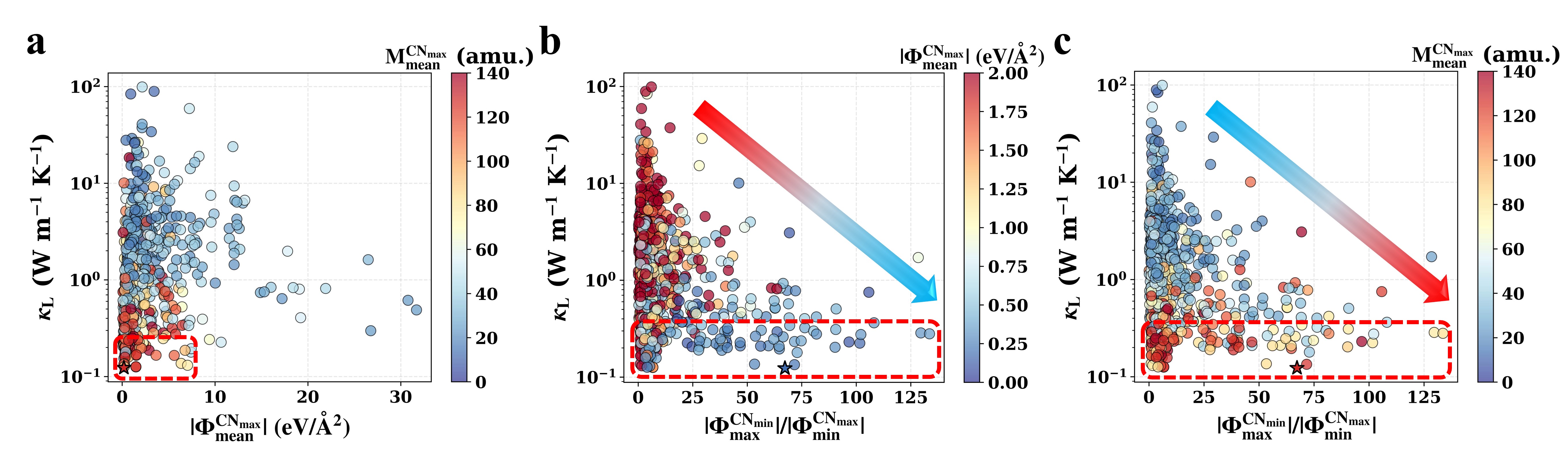}% Here is how to import EPS art
\caption{\label{fig:5} \textbf{Dependence of $\kappa_{\rm L}$ on structural IFC descriptors for materials reported in this work.} \textbf{a}, $\kappa_{\rm L}$ as a function of the average harmonic IFCs of the polyhedron with the maximum coordination number ($\abs{\rm \Phi^{CN_{max}}_{mean}}$). \textbf{b-c}, $\kappa_{\rm L}$ versus the ratio between the maximum and minimum harmonic IFCs for polyhedra with the minimum and maximum coordination numbers, respectively ($\abs{\rm \Phi^{CN_{min}}_{max}}/\abs{\rm \Phi^{CN_{max}}_{min}}$). The colorbars represent the mean atomic mass of the maximum-coordination polyhedron ($\rm M^{CN_{max}}_{mean}$) in panels \textbf{(a)} and \textbf{(c)}, and the $\abs{\rm \Phi^{CN_{max}}_{mean}}$ in panel \textbf{(b)}. The asterisk denotes CsTlI$_4$ identified in this work.}
\end{figure*}

Understanding the ultralow $\kappa_{\rm L}$ of CsTlI$_4$ requires an in-depth analysis of its bonding environment and vibrational properties. Figure \ref{fig:4}a presents the harmonic interatomic force constants (IFCs) of nearest-neighbor atom pairs in CsTlI$_4$, benchmarked against a series of ultralow-$\kappa$ materials. The Cs-I interactions are exceptionally weak, nearly 60 times weaker than the Tl-I bonds within the same structure. This large disparity in stiffness, more pronounced than that found in other compounds, is closely associated with the presence of a wide phonon frequency gap. The softness of Cs–I bonding is mainly due to the high coordination number and the formation of low-symmetry polyhedra around Cs atoms. Notably, the materials identified in our high-throughput screening (Table S1) and many other reported ultralow-$\kappa_{\rm L}$ compounds predominantly involve alkali metal–halogen combinations. To place this bonding feature in a broader context, we analyzed a series of alkali halides, as shown in Fig. S9. Weaker $A$–$X$ interactions are generally associated with lower $\kappa_{\rm L}$, with Cs–I exhibiting the smallest interatomic force constants and the lowest $\kappa_{\rm L}$ within this series. This trend suggests that weak bonding in such systems can contribute to suppressed heat transport.

To probe the generality of this mechanism, we evaluated the thermal transport behavior across a family of Cs-I compounds, including CsI (Fm$\bar{3}$m), CsI (Pm$\bar{3}$m), CsI$_3$, and CsI$_4$, as illustrated in Figs. \ref{fig:4}b-e and S10-12. The $\kappa_{\rm L}$ decreases consistently as the Cs coordination number (CN) increases and the polyhedral symmetry decreases. As shown in Fig. \ref{fig:4}c, higher coordination leads to weaker and more widely distributed interatomic force constants, reflecting reduced bond stiffness (Fig. S12) and enhanced bond hierarchy. This weak bonding results in enhanced anharmonicity and reduced phonon group velocities (Fig. S11). In addition, the broadened distribution of bond strengths gives rise to a separation of vibrational frequencies, where Cs atoms dominate low-frequency modes while I atoms contribute primarily at higher frequencies (Fig. \ref{fig:4}d). Such a combined effect of strong anharmonicity, reduced phonon group velocities, and vibrational mismatch suppresses both particle-like phonon propagation ($\kappa^{\rm p}_{\rm L}$) and wave-like coherent transport ($\kappa^{\rm c}_{\rm L}$), as evidenced in Fig. \ref{fig:4}e.

Further insight into the origin of the distinct bonding hierarchy in CsTlI$_4$ is obtained from crystal orbital Hamilton population (-COHP) analysis (Fig. \ref{fig:4}f). The results reveal that Tl–I interactions exhibit pronounced bonding states below the Fermi level, indicating strong interatomic interactions, while the presence of antibonding states near the Fermi level suggests a tendency toward lattice softening. In contrast, Cs-I bonds are very weak, suggesting a structurally soft Cs-centered sublattice. The coexistence of these two sublattices introduces a clear hierarchy in bonding strength. The charge density map in Fig. \ref{fig:4}g visually highlights the bonding hierarchy that Tl–I bonds exhibit electron density overlap, while Cs–I regions lack clear bonding features. This contrast is further supported by reduced density gradient (RDG) analysis in Fig. \ref{fig:4}h-i, which indicates that Cs–I interactions are governed by weakly dispersive van der Waals forces, whereas Tl–I pairs are characterized by much stronger interatomic interactions. Collectively, these results establish the presence of hierarchical bonding in CsTlI$_4$, where weak and low-symmetry Cs-I networks coexist with more robust Tl-I sublattices. This hierarchical structure leads to a vibrational-mode mismatch between the Cs- and Tl-based sublattices, which further disrupts phonon coherence and propagation, providing a microscopic explanation for the observed ultralow $\kappa_{\rm L}$.

Building upon the bonding hierarchy identified in Fig. \ref{fig:4}, we further examine how bonding descriptors correlate with $\kappa_{\rm L}$ across a broader set of compounds. As shown in Fig. \ref{fig:5}a, the $\kappa_{\rm L}$ exhibits a clear dependence on the harmonic interatomic force constants (IFCs) associated with the polyhedron possessing the maximum coordination number ($\abs{\rm \Phi^{CN_{max}}_{mean}}$). Materials with smaller average IFCs tend to display lower $\kappa_{\rm L}$, particularly when the polyhedron contains heavier atoms. This trend is consistent with the fundamental relationship governing phonon group velocity, where $v_g \propto (k/\rm M)^{1/2}$ with $k$ representing bond stiffness and $\rm M$ the atomic mass. Weak bonding not only reduces phonon group velocities but also enhances anharmonicity, both of which contribute to the suppression of thermal transport. The clustering of ultralow-$\kappa_{\rm L}$ materials within the region of small IFCs and large atomic mass further highlights that low group velocity and strong anharmonicity are dominant prerequisites for achieving ultralow $\kappa_{\rm L}$.

Beyond bond softness, the mismatch in bonding strength between coordination environments also plays a critical role. Figures \ref{fig:5}b and \ref{fig:5}c reveal that the $\kappa_{\rm L}$ systematically decreases as the ratio between the maximum and minimum harmonic IFCs ($\abs{\rm \Phi^{CN_{min}}_{max}}/\abs{\rm \Phi^{CN_{max}}_{min}}$) increases. This ratio characterizes the degree of IFC mismatch between the polyhedra with the minimum and maximum coordination numbers. A larger mismatch corresponds to a stronger bonding hierarchy within the crystal structure, which induces a separation of vibrational frequencies across sublattices and disrupts vibrational continuity. As the IFC mismatch increases, materials show a progressively stronger tendency toward low $\kappa_{\rm L}$, indicating that bonding hierarchy serves as an effective structural descriptor for identifying crystalline thermal insulators. Notably, compounds located within the dashed low-$\kappa_{\rm L}$ region typically exhibit weak bonding in highly coordinated polyhedra, often combined with large atomic mass, which leads to reduced phonon group velocities and enhanced anharmonicity. At the same time, a pronounced bonding hierarchy introduces vibrational frequency mismatch across sublattices. These effects act cooperatively to suppress thermal transport by simultaneously reducing phonon propagation and disrupting vibrational coherence. This unified picture explains the extremely low $\kappa_{\rm L}$ of CsTlI$_4$ and provides physically interpretable descriptors for identifying ultralow-$\kappa_{\rm L}$ materials.

\section{Discussion}

In this work, we developed a high-throughput computational screening framework that integrates universal machine-learning interatomic potentials with temperature-renormalized lattice dynamics, higher-order anharmonic scattering, and unified thermal transport theory for accurate prediction of lattice thermal conductivity. Using this framework, we identified CsTlI$_4$ as a candidate crystalline thermal insulator and experimentally confirmed its record-low thermal conductivity of 0.14 $\rm W m^{-1} K^{-1}$ at 300 K, representing the lowest value reported for a dense single-phase crystalline solid without superlattice interfaces.

Detailed lattice dynamics and bonding analyses reveal that the ultralow $\kappa_{\rm L}$ originates from a pronounced hierarchical bonding structure. Weakly bonded Cs–I polyhedra soften the lattice and significantly reduce phonon group velocities, while stronger Tl–I interactions introduce large disparities in bonding stiffness between coordination environments. This bonding hierarchy gives rise to a vibrational mismatch between sublattices and disrupts vibrational continuity. As a result, both particle-like phonon propagation and wave-like interband tunneling contributions to thermal transport are strongly suppressed.

More broadly, correlations identified between lattice thermal conductivity and structural descriptors derived from interatomic force constants suggest two general design principles for crystalline thermal insulators. Weak bonding in highly coordinated polyhedra, particularly those containing heavy elements, and a strong bonding hierarchy between coordination environments act cooperatively to suppress heat transport by reducing phonon velocities, enhancing anharmonicity, and inducing vibrational mismatch. The high-throughput framework and physically interpretable descriptors established in this work provide a pathway for transforming the discovery of thermal insulating materials from empirical exploration to predictive and data-driven design.

\section{Methods}

\paragraph{\textbf{High-throughput identification of weakly bonded semiconductors.}}
Crystal structures were retrieved from the Materials Project (MP) dataset \cite{jain2013commentary} through the MP API. To define a chemically simple and computationally tractable search space for high-throughput lattice thermal conductivity calculations, a series of filters were applied prior to structural analysis. Compounds satisfying the following criteria were retained: energy above the convex hull $E_\mathrm{hull} \leq 0.05$ eV atom$^{-1}$, band gap between 0.1 and 3.0 eV, and no more than 25 atomic sites in the primitive cell. In addition, only binary and ternary compounds were considered, and all structures containing rare-earth elements were excluded. Metallic and magnetic entries were removed according to the MP metadata, including the \texttt{is\_metal} flag, total magnetization, and reported magnetic ordering.

Weakly bonded materials were identified from their local coordination environments using the CrystalNN algorithm \cite{pan2021benchmarking} as implemented in pymatgen \cite{ong2013python}. For each atomic site $i$, the first coordination shell was constructed with a search cutoff of 5 \AA. The interatomic distance $D_{ij}$ between atom $i$ and a nearest neighbor $j$ was then compared with the sum of their tabulated covalent radii, $R_i + R_j$, using the covalent radii reported by Cordero \textit{et al.} \cite{cordero2008covalent}. Following the descriptor introduced by Li \textit{et al.}\cite{jacs1c11887}, a neighboring pair was classified as weakly bonded when $D_{ij} > F_1 (R_i + R_j)$, where $F_1$ is a dimensionless scaling factor. In this work, $F_1 = 1.15$ was adopted to identify elongated bonds while retaining chemically reasonable structures.

\paragraph{\textbf{Initial screening of lattice thermal conductivity using a pretrained Orb potential.}}
For each candidate material, the primitive cell was expanded into an approximately isotropic supercell containing at least 100 atoms to represent thermal phase space and anharmonic interactions with sufficient configurational freedom while minimizing artificial shape anisotropy. Molecular dynamics (MD) simulations were then carried out at 300 K using the conservative Orb-v3 universal interatomic potential \cite{rhodes2025orb} within the Atomic Simulation Environment (ASE) \cite{hjorth2017atomic}. Initial atomic velocities were assigned according to the Maxwell-Boltzmann distribution, and canonical sampling was performed using a Langevin thermostat with a time step of 1 fs. For each structure, 25,000 MD steps were performed, from which 200 thermally displaced configurations were extracted from the equilibrated portion of the trajectory at uniform time intervals.

The total energies and atomic forces of these displaced configurations were evaluated using the same pretrained Orb-v3 potential and used as input to fit high-order interatomic force constants (IFCs) within the Hiphive framework \cite{eriksson2019hiphive}.  The cutoff for the third- and fourth-order IFCs were set to the fourth and second nearest-neighbor shells, respectively. Finite-temperature anharmonic renormalization was incorporated through self-consistent phonon (SCPH) \cite{PRB92054301} calculations at 300 K using Hiphive. Both the SCPH iterations and fitting structures were set to 100, and a linear mixing parameter of 0.1 was employed to stabilize convergence. Phonon dispersions were calculated using Phonopy \cite{togo2023first}, and structures exhibiting imaginary phonon branches were classified as dynamically unstable at 300 K and excluded from subsequent calculations.

Three-phonon scattering was treated iteratively, while four-phonon scattering was evaluated within the relaxation time approximation (RTA) using the FourPhonon code \cite{HAN2022108179, guo2024sampling}.  The reciprocal-space sampling mesh was chosen adaptively according to the primitive-cell size and average lattice parameters to balance numerical accuracy against computational cost. Energy conservation was treated using a Gaussian broadening of 0.1, and $10^6$ four-phonon scattering events were sampled to ensure statistical convergence. The lattice thermal conductivity $\kappa_{\rm L}$ at 300 K was calculated within the unified thermal transport framework that includes both particle-like ($\kappa_{\rm L}^{\rm p}$) and wave-like contributions ($\kappa_{\rm L}^{\rm c}$) \cite{UT2019}. 

\paragraph{\textbf{Accurate evaluation of lattice thermal conductivity through material-specific fine-tuning.}}
For materials identified in the initial screening stage with $\kappa_\mathrm{L} \leq 0.2$ $\rm W m^{-1} K^{-1}$, a refined workflow was employed to improve the quantitative accuracy $\kappa_{\rm L}$. Each structure was first fully relaxed. The relaxed primitive cell was then expanded into a supercell containing at least 100 atoms, and MD simulations at 300 K were performed using the same protocol as described above. From each trajectory, 200 thermally displaced configurations were extracted. For each material, 2\% of these configurations were randomly selected for first-principles self-consistent calculations. These labeled configurations were subsequently used to fine-tune the pretrained conservative Orb-v3 model \cite{rhodes2025orb}.

Fine-tuning was performed using an ASE-compatible SQLite dataset with total energies and atomic forces as the training targets. Random rotational augmentation was applied during training to improve generalization with respect to structural orientation. The dataset was randomly split into training and test subsets with an 8:2 ratio. Full-parameter fine-tuning was performed to fully adapt the pretrained model to the local potential-energy surfaces of the target low-$\kappa_{\rm L}$ materials. Model optimization employed the default Orb loss function, which jointly accounts for energy and force errors. In the conservative formulation, atomic forces are obtained as analytical energy gradients with respect to atomic positions.

Training was conducted on a single GPU using a step-based schedule with 64 steps per epoch for up to 200 epochs. The learning rate was set to $1 \times 10^{-5}$, and gradient clipping with a maximum norm of 0.5 was applied. A batch size of 4 was used. Model checkpoints were saved during training, and the best-performing checkpoint was used to predict the energies and forces of the remaining displaced configurations for each material. The combined dataset of labeled configurations and fine-tuned predictions was then used to refit the IFCs. The lattice thermal conductivity was subsequently recomputed following the same workflow used in the initial screening stage.

\paragraph{\textbf{Density functional theory calculations.}} 
Density functional theory (DFT) calculations were performed using the Vienna \textit{Ab initio} Simulation Package (VASP) \cite{kresse1996VASP}, employing projector-augmented wave (PAW) pseudopotentials \cite{PRB5017953} and the PBEsol functional \cite{PBEsol2008} within the generalized gradient approximation (GGA) \cite{1996PBE}. The valence configurations considered were Cs (5$s^2$5$p^6$6$s^1$), Tl (5$d^{10}$6$s^2$6$p^1$), and I (5$s^2$5$p^5$). A plane-wave energy cutoff of 520 eV and a $\rm \Gamma$-centered Monkhorst-Pack \textit{k}-point mesh with a spacing of approximately $2\pi \times 0.025$ \AA$^{-1}$ were adopted for Brillouin zone integration. All structures were fully relaxed until the residual atomic forces were below 10$^{-4}$ eV/\AA\ and the total energy converged to 10$^{-8}$ eV. The optimized lattice parameters are listed in Table S8 and are consistent with experimental data. To evaluate the anharmonicity parameter $\sigma$ \cite{PRM4083809} of CsTlI$_4$, \textit{ab initio} molecular dynamics (AIMD) simulations were conducted using $3 \times 1 \times 3$ supercells (216 atoms) and a 400 eV cutoff. A $\rm \Gamma$-centered $1 \times 1 \times 1$ $k$-mesh was used with a time step of 2 fs and 10,000 steps.

\paragraph{\textbf{Extraction of interatomic force constants.}}
To extract accurate temperature-dependent interatomic force constants (IFCs), we performed \textit{ab initio} molecular dynamics (AIMD) simulations from which 60 configurations were randomly selected at each temperature, and static DFT calculations were conducted to obtain corresponding atomic forces. These force-displacement datasets were then used to fit second-order IFCs within the self-consistent phonon (SCPH) framework \cite{PRB1572, SCPH2014} as implemented in the Hiphive package \cite{eriksson2019hiphive}, while the harmonic IFCs at 0 K were independently calculated using the finite displacement method in Phonopy \cite{TOGO20151} with a displacement amplitude of 0.01 \AA. To accurately capture the anharmonic effects, third- and fourth-order IFCs were obtained using Hiphive by subtracting harmonic forces prior to cluster space training, ensuring that only anharmonic force components were retained in the fitting process. Based on convergence tests, $3 \times 1 \times 3$ supercells of CsTlI$_4$ were employed, and the cutoff radii for the second-, third-, and fourth-order IFCs were set to 9.0, 6.0, and 4.5 \AA, respectively (Fig. S4). For comparative analysis, we further evaluated a series of cesium iodine compounds using supercells and cutoff parameters to accurately represent their harmonic and anharmonic interatomic interactions: CsI in the Fm$\bar{3}$m and Pm$\bar{3}$m phases were modeled using $4 \times 4 \times 4$ supercells. The cutoff radii for the second-, third-, and fourth-order interatomic force constants (IFCs) were set to 9.0, 7.0, and 6.0 \AA\ for the Fm$\bar{3}$m phase, and 8.0, 7.0, and 6.0 \AA\ for the Pm$\bar{3}$m phase, respectively. For CsI$_3$ and CsI$_4$, $2 \times 3 \times 2$ and $2 \times 2 \times 2$ supercells were employed, with corresponding second-, third-, and fourth-order IFC cutoff radii of 9.0, 6.5, and 5.0 \AA, and 8.0, 6.0, and 4.5 \AA, respectively.

\paragraph{\textbf{Synthesis of CsTlI$_4$.}} 
Cesium iodine (CsI, 99.999\%), iodine (I$_2$, 99.99\%), and thallium iodine (TlI, 99.99\%) were purchased from Aladdin (China), while bismuth (III) iodine (BiI$_3$, 99\%) was bought from Heowns (China). All chemicals were used as received without further purification.

For CsTlI$_4$, CsI (1 mmol), TlI (1 mmol), and I$_2$ (1 mmol) were thoroughly ground under an argon atmosphere in a glovebox. The mixed powder was sealed in a quartz tube under vacuum (10$^{-4}$ Pa) and heated to 723 K in a muffle furnace at 10 K min$^{-1}$, held for 2 h, followed by cooling to 493 K at the same rate. The temperature was maintained at 493 K for 48 h before natural cooling to room temperature. The sintered powder was cold-pressed into a pellet (11 mm in diameter and 1.5 mm in thickness), resealed in a quartz tube under vacuum (10$^{-4}$ Pa), and further sintered at 423 K for 24 h. The pellet density was measured by the Archimedes method and reached 97\% of the theoretical value (5.02 g cm$^{-3}$).

\paragraph{\textbf{Synthesis of Cs$_3$Bi$_2$I$_9$.}}
For Cs$_3$Bi$_2$I$_9$, stoichiometric amounts of 3 mmol of CsI and 2 mmol of BiI$_3$ were thoroughly ground in a glovebox under argon atmosphere. The mixture was loaded into a quartz ampoule, which was evacuated (10$^{-4}$ Pa) and flame-sealed. The ampoule was heated to 1023 K at 3 K min$^{-1}$, and held for 24 h, then cooled to room temperature over 48 h at 2 K min$^{-1}$. The resulting ingots were ground into fine powders and consolidated by spark plasma sintering (SPS) at 593 K under a pressure of 17 MPa for 5 min in high vacuum. The final pellet density reached 98\% of the theoretical value (4.76 g cm$^{-3}$).

\paragraph{\textbf{Synthesis of CsI.}}
A total of 1 mmol of CsI was ground in an argon atmosphere inside a glovebox, followed by cold-pressing into a circular pellet (11 mm diameter and 1.8 mm thickness) and sealing in a quartz tube under a vacuum of 10$^{-4}$ Pa. The sealed tube was subsequently sintered at 473 K for 24 h. The density of the resulting pellet was measured using the Archimedes method and found to be 96\% of the theoretical value (4.51 g cm$^{-3}$).

\paragraph{\textbf{Synthesis of CsI$_3$.}}
For CsI$_3$, 1 mmol of CsI and 1 mmol of I$_2$ were loaded into a zirconia milling jar under an argon atmosphere. Ball milling was carried out using a YXQM-8L planetary high-energy mill (Changsha MITR Instruments, China) at 800 rpm for 8 h, with a powder-to-ball weight ratio of 1:10 and zirconia balls of mixed diameters (10, 8, and 5 mm). The obtained powder was cold-pressed into a pellet (11 mm in diameter and 1.8 mm in thickness), sealed in a quartz tube under vacuum (10$^{-4}$ Pa), and sintered at 353 K for 10 h with a heating rate of 10 K min$^{-1}$. The pellet density was determined to be 96\% of the theoretical value (4.48 g cm$^{-3}$).

\paragraph{\textbf{XRD measurement.}}
The crystal structures were characterized using a Haoyuan DX-27MINI powder X-ray diffractometer with Cu K$\alpha_1$ radiation ($\lambda = 1.5406$ \AA). Diffraction patterns were collected over a 2$\rm \theta$ range of 10–80$\degree$ at room temperature, with an operating voltage of 40 kV and a current of 40 mA. Rietveld refinements were carried out using the FullProf Suite \cite{RODRIGUEZCARVAJAL199355} and WinPLOTR \cite{roisnel2001winplotr} programs, during which the background, zero shift, lattice parameters, profile parameters, isotropic displacement parameters ($U_{\rm iso}$), and atomic positions were refined. The refined structural parameters are summarized in Tables S3–6.

\paragraph{\textbf{Thermal conductivity measurements.}} 
The lattice thermal conductivity was estimated via the formula $\kappa_{\rm L}$ = $\rho$$C_p$$d$. Thermal diffusivity (\textit{d}) of the Cs-based compounds was measured using a Netzsch LFA 467 Hyperflash system under a nitrogen atmosphere over temperature ranges of 163–423 K for CsTlI$_4$, 173–300 K for Cs$_3$Bi$_2$I$_9$, 173–573 K for CsI, and 163–300 K for CsI$_3$. The density ($\rho$) was obtained via the Archimedes method, and specific heat capacity ($C_p$) was determined using Dulong-Petit law.

\paragraph{\textbf{Sound velocity measurements.}}
The transverse ($v_{\rm T}$) and longitudinal ($v_{\rm L}$) sound velocities (Table S7) of the CsTlI$_4$ sample with a diameter of 10 mm and thickness of 2 mm were recorded at 300 K using an ultrasonic measurement system. A small amount of coupling grease was applied between the transducers and the sample surfaces to improve acoustic coupling. The average sound velocity $v_{\rm ave}$ of the sample was calculated from the measured $v_{\rm L}$ and $v_{\rm T}$ using the following relation \cite{PRB94125203} : $v_{\rm ave}=[\frac{1}{3}(\frac{1}{{v_{\rm L}}^{3}}+\frac{2}{{v_{\rm T}}^{3}})]^{-1/3}$.

\textbf{Data availability}
The data supporting the findings of this study are available at Supplementary Forms I and II.

\textbf{Code availability}
The universal machine-learning interatomic potential (uMLIP) used in this work is based on the Orb-v3 model \cite{rhodes2025orb}. Density functional theory (DFT) calculations and \textit{ab initio} molecular dynamics (AIMD) simulations were carried out using the VASP package \cite{kresse1996VASP, 1999PAW}. Interatomic force constants were constructed with the Hiphive package \cite{eriksson2019hiphive}. Phonon linewidths and lattice thermal conductivity were calculated using the FourPhonon \cite{HAN2022108179} and Phono3py \cite{PRB91094306, UT2019} packages, respectively, with the latter extended in-house to incorporate four-phonon scattering processes. Chemical bonding was characterized via crystal orbital Hamilton population (COHP) analysis as implemented in the LOBSTER package \cite{j100135a014, jp202489s}. Non-covalent interaction (NCI) analysis was performed using the Critic2 package \cite{OTERODELAROZA20141007}, and the corresponding visualization was carried out with the visual molecular dynamics (VMD) package \cite{HUMPHREY199633}. Machine learning neuroevolution potentials (NEP) were trained, and homogeneous nonequilibrium molecular dynamics (HNEMD) simulations were performed using the GPUMD package \cite{FAN201710}. 

\section{References}
%\nocite{*}
\bibliography{HTC_CsTlI4}

\par
\textbf{Acknowledgments}
This work is supported by the Research Grants Council of Hong Kong (C7002-22Y and C1002-24Y). X.S. acknowledges financial support from the National Youth Talent Project (Overseas, D5113250165) and the Fundamental Research Funds for the Central Universities (D5000250021, GH20260201339). X.S. and E.G. acknowledge funding from the European Union’s Horizon 2020 Research and Innovation Programme under the Marie Sklodowska-Curie Grant Agreement No. 101034329 and the WINNING Normandy Programme supported by the Normandy Region. C.W. acknowledges the National Natural Science Foundation of China (No. 12504019) and the Guangdong Pearl River Talent Program (No. ZJQNRC20241219163147045). P.L. and J.H. acknowledge the Operational Program Research, Development and Education financed by the European Structural and Investment Funds and by the Ministry of Education, Youth and Sports (MEYS) of the Czech Republic, Grant No. CZ.02.01.01/00/22 008/0004594 (TERAFIT). The authors are grateful for the research computing facilities offered by the ITS, HKU.

\textbf{Author Contributions Statement}
R.C. and Y.C. conceived the idea and designed the project. R.C., Z.C., C.W., and Z.Z. wrote the code and performed the calculations. M.J., L.J., P.L., J.H., C.C., E.G., and X.S. performed the experiments. All authors contributed to the results discussion and paper writing.

\textbf{Competing Interests Statement}
The authors declare no competing interests.

\end{document}